\documentclass[12pt]{article}
\usepackage{epsfig}

\def\vec#1{\mbox{\boldmath$#1$}}
\def\({\left(} \def\){\right)}  
\def\lk{\,\left[ \,} \def\rk{\,\right] \,}
   
\def\be{ \begin{equation} }  \def\bea{ \begin{eqnarray} }
\def\ee{ \end{equation} }    \def\eea{ \end{eqnarray} }
\def\Av#1{\left\langle \vphantom{A^b_\mu} #1 \right\rangle}
\newcommand{\msb}{{\overline{{\rm MS}}}}
\newcommand{\msbb}{{{}\over{\rm MS}}}

\let\para=\S

\def\b{\beta}

\def\d{\delta}
\def\e{\epsilon}

\def\g{\gamma}

\def\m{\mu}

\def\p{\pi}

\def\x{\xi}
\def\z{\zeta}

\def\S{\Sigma}

\def\ve{\varepsilon}

\def\pl#1#2#3{Phys.~Lett.~{\bf B {#1}} (19{#2}) #3}
\def\np#1#2#3{Nucl.~Phys.~{\bf B {#1}} (19{#2}) #3}
\def\prl#1#2#3{Phys.~Rev.~Lett.~{\bf #1} (19{#2}) #3}
\def\pr#1#2#3{Phys.~Rev.~{\bf D {#1}} (19{#2}) #3}

\def\prep#1#2#3{Phys.~Rep.~{\bf {#1}C} (19{#2}) #3}

\begin{document}
\date{\mbox{ }}

\title{ 
{\normalsize     
DESY 98-191 \hfill\mbox{}\\
hep-ph/9812205 \hfill\mbox{}\\
December 1998 \hfill\mbox{}\\}
\vspace{2cm}
\bf THE STATIC POTENTIAL IN QCD\\ TO TWO LOOPS\\[8mm]}
\author{York~Schr\"oder~\footnote{e--mail: {\tt York.Schroeder@desy.de}}\\
{\it Deutsches Elektronen-Synchrotron DESY, 22603 Hamburg, Germany}}
\maketitle

\thispagestyle{empty}

\vspace{1cm}
\begin{abstract}
\noindent
We evaluate the static QCD potential to two--loop order.
Compared to a previous calculation a sizable reduction
of the two--loop coefficient $a_2$ is found.
\end{abstract}

\newpage

\section{Introduction}

The static potential of (massless) QCD has recently been 
calculated to two loops \cite{Pe}. 
Being a fundamental quantity, it is of importance
in many areas, such as NRQCD, quarkonia, quark mass definitions and
quark production at threshold.
While the one--loop contribution and the two--loop
pole terms have been known for a long time \cite{Su,Fi,Bi},
the two--loop constant $a_2$ (cf.\ Eqs.~(\ref{Result})-(\ref{resulT}))
was found only recently
\cite{Pe}. The fermionic parts of this coefficient
were confirmed numerically in \cite{Me}, taking the $m_q\!\rightarrow\!0$
limit of a calculation involving massive fermion loops. 
The aim of this work is to evaluate $a_2$ analytically, 
using an independent method. 

The static potential is defined in a manifestly
gauge invariant way via the vacuum expectation
value of a Wilson loop~\cite{Bi,mgi_def},
\be \label{def_WL}
V(r) = - \lim_{T\rightarrow\infty} \frac1T\, \ln 
\Av{ {\rm tr} {\cal P} \exp\(ig \oint_\Gamma dx_\mu A_\mu\) } \;.
\ee
Here, $\Gamma$ is taken as a rectangular loop with time extension 
$T$ and spatial extension $r$, and $A_\m$ is the vector potential 
in the fundamental representation.

In a perturbative analysis it can be shown that, 
at least to the order needed here, all 
contributions to Eq.~(\ref{def_WL}) containing connections to
the spatial components of the gauge fields $A_i(\vec r,\pm T/2)$
vanish in the limit of large time extension $T$.
Hence, the definition can be reduced to 
\be \label{def2}
V_{\rm pert}(r) = - \lim_{T\rightarrow\infty} \frac1T\, \ln
\Av{ {\rm tr} {\cal T} \exp\(-\int_x J_\mu^a A_\mu^a\) } \;,
\ee
where ${\cal T}$ means time ordering and the static sources 
separated by the distance $r=|\vec r-\vec r\,'|$ are given by
\be
J_\mu^a(x) = ig \, \delta_{\mu 0} T^a \lk \delta(\vec x-\vec r) 
-\delta(\vec x-\vec r\,') \rk \;,
\ee
where $T^a$ are the generators in the fundamental representation. 
In the case of QCD the gauge group is $SU(3)$. The calculation 
will be carried out for an arbitrary compact semi-simple Lie group 
with structure constants
defined by the Lie algebra $[T^a,T^b]=if^{abc}T^c$. 
The Casimir operators of the fundamental and adjoint representation
are $T^aT^a=C_F$ and $f^{acd}f^{bcd}=C_A\delta^{ab}$. 
$tr(T^aT^b)=T_F\delta^{ab}$ is the trace normalization, 
while $n_f$ denotes the number of massless quarks. 

Expanding the expression in Eq.~(\ref{def2}) perturbatively, one 
encounters in addition to the usual Feynman rules the source--gluon 
vertex $ig \delta_{\mu 0} T^a$, with an additional minus sign for the
antisource. Furthermore, the time--ordering prescription
generates step functions, which can be viewed as source
propagators, analogous to the heavy--quark effective theory
(HQET)~\cite{HQET}.  

\begin{figure}
\begin{center}
\psfig{file=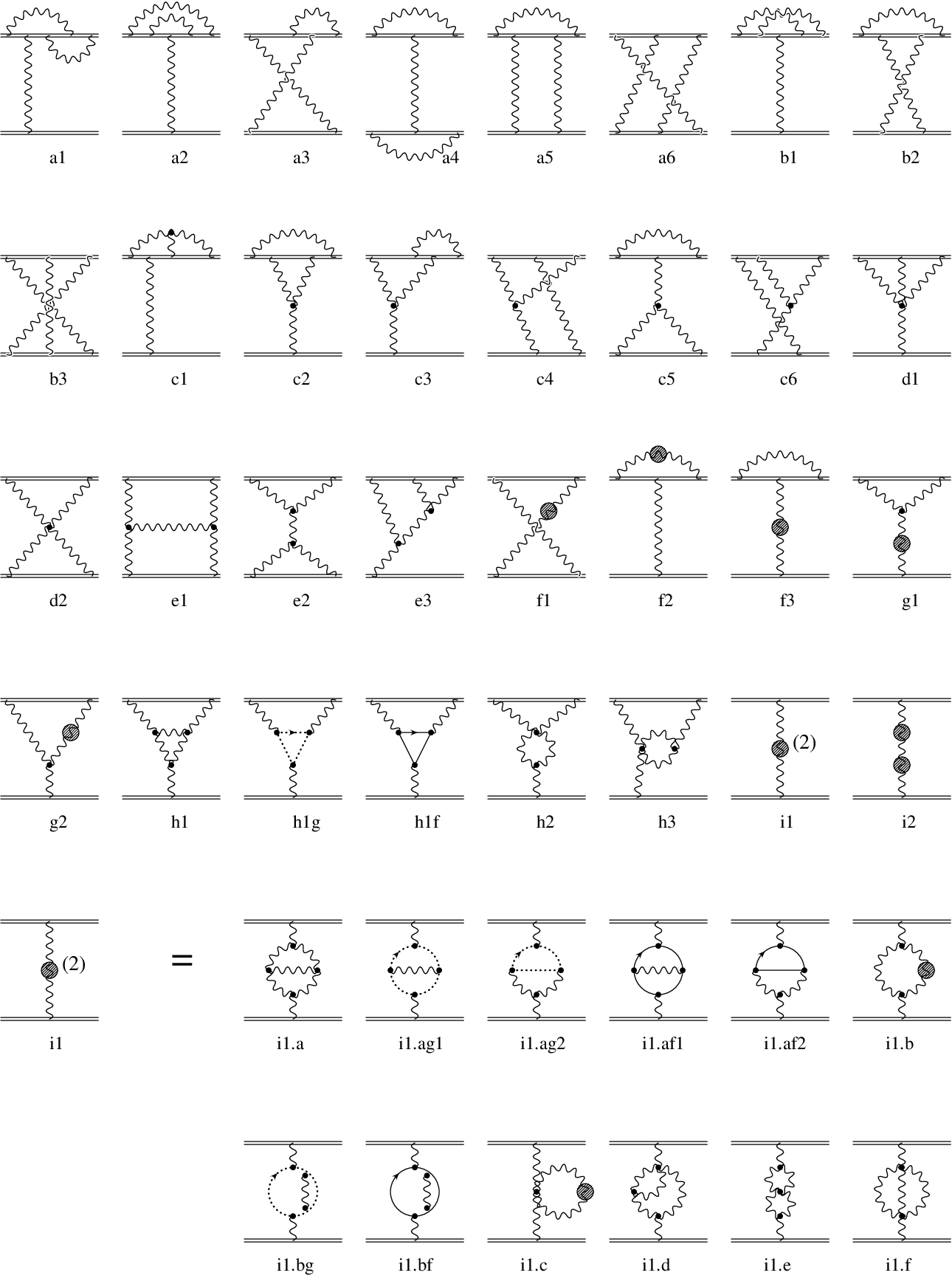,width=12cm}
\end{center}
\caption{\label{fig1} Classes of two--loop diagrams 
contributing to the static potential.
Double, wiggly, dotted and solid lines denote source, gluon, ghost 
and (light) fermion propagators, respectively. A blob on a 
gluon line stands for one--loop self--energy corrections. }
\end{figure}

Concerning the generation of the complete set of Feynman diagrams
contributing to the two--loop static potential, there are some
subtleties connected with the logarithm in the 
definition~(\ref{def2}).
All this is explained in detail in \cite{Pe,Fi}, so we
only list the relevant diagrams here (see Fig.\ref{fig1}). Note that
the aforementioned papers are based on the Feynman gauge, while
we use general covariant gauges, resulting in an enlarged set
of diagrams.

\section{Method}

The method employed in this work can be briefly 
summarized as follows:

$\bullet$ All dimensionally regularized (tensor-) integrals 
are reduced to pure propagator
integrals by a generalization of Tarasov's method \cite{Ta}.
The resulting expressions are then mapped to
a minimal set of five scalar integrals by means
of recurrence relations, again generalizing \cite{Ta}
as well as \cite{ChTk} to the case including
static (noncovariant) propagators\footnote{This algorithm 
will be described in detail elsewhere \cite{YSprep}.}. 
These two steps are implemented into a FORM \cite{Form} package.
Thus, we constructed our method to be complementary to 
the calculation in \cite{Pe}, assuring a truly independent
check. At this stage, one obtains analytic coefficient functions 
(depending on the generic space--time dimension $D$ as well
as on the color factors and the bare coupling), multiplying 
each of the basic integrals.

$\bullet$ The basic scalar integrals are then solved analytically.
Expanding the result around $D=4$ (which is done in both
MAPLE \cite{Maple} and Mathematica \cite{Mathematica} 
considering the complexity
of the expressions) and renormalizing, one obtains
the final result to be compared with \cite{Pe}.
 
Important checks of the calculation are the comparison of
the pole terms of individual gluonic diagrams given in \cite{Fi}, 
the gauge independence of appropriate classes of diagrams, 
the confirmation of cancellation of infrared divergencies, 
and the correct renormalization properties.

\section{Renormalization and result}

The renormalized quantities
in $D=4-\e$ dimensions are conventionally defined by 
\be V \equiv \m^\e V_R \quad,\quad
    \frac{g^2}{16 \p^2} \equiv Z \m^\e a_R \;,
\ee
where the subscript {\footnotesize $R$} denotes the renormalized
quantities.
The factor $Z$ is assumed to have an expansion
in the renormalized coupling, 
$Z = 1 + a_R Z^{'}(\e) + a_R^2 Z^{''}(\e) + ...$,
and we choose to work in the $\msb$ scheme, related to the MS scheme
by the scale redefinition
$\m^2 = \bar\m^2 e^\g/4\p$ .

The needed counterterms read explicitly
\be \label{zets}
  Z^{'}_{\msbb} = -\frac2\e\,\b_0 \quad,\quad 
  Z^{''}_{\msbb} = \frac4{\e^2}\,\b_0^2 
  - \frac1\e\,\b_1 \;.
\ee
Here, the coefficients of the Beta function 
are defined by the running coupling, 
$\m^2\partial_{\m^2}\,a_R = -\b_0a_R^2-\b_1a_R^3-...$,
$\b_0 = \frac{11}3\,C_A - \frac43\,T_Fn_f$
and
$\b_1 = \frac{34}3\,C_A^2 - 4C_FT_Fn_f -\frac{20}3\,C_AT_Fn_f$.
The results of our calculation yield indeed Eq.~(\ref{zets}).

As a further check on the pole terms, the vertex and gluon
wave function renormalization constants ($Z_1$ and $Z_3^{-1}$,
respectively)
have been extracted separately from the diagrams. They depend
on the gauge parameter $\x$ and agree with the ones
given in \cite{ren}.

The renormalized potential now reads
$V_{\rm pert}(r) = \int\!\frac{d^3q}{(2\p)^3}\, 
\exp(i\vec q\vec r) V(\vec q^2)\,$,
with
\newcommand{\mq}{\left(\frac{\bar\m^2}{\vec q^2}\right)}
\bea \label{Result}
  V(\vec q^2) &=& - \frac{C_F 16\p^2}{\vec q^2}\, a_\msbb 
  \left\{ 1 + a_\msbb\, c_{1\,\msbb}\mq
  + a_\msbb^2\, c_{2\,\msbb}\mq + ... \right\} 
\eea
where
\bea 
  c_{1\,\msbb}(x) &=& a_1 + \b_0 \ln(x) \;,\\
  c_{2\,\msbb}(x) &=& a_2 + \b_0^2\ln^2(x) + (\b_1+2\b_0a_1)\ln(x) 
\eea
and
\bea
  a_1 &=& \frac{31}9\,C_A - \frac{20}9\,T_Fn_f \;, \\
  a_2 &=& \left( \frac{4343}{162} + 4\p^2 - \frac{\p^4}4 + 
  \frac{22}3 \z(3) \right) C_A^2 - \left( \frac{1798}{81} + 
  \frac{56}3\,\z(3) \right) C_AT_Fn_f \nonumber\\
  &&- \left( \frac{55}3 - 16\z(3)
  \right) C_FT_Fn_f + \frac{400}{81}\,T_F^2n_f^2 \;. \label{resulT}
\eea
As it has to be, the coefficients prove to be gauge  
independent.
Comparing our two-loop result for $a_2$ with \cite{Pe}, we find
a discrepancy of $2\p^2$ in the pure Yang--Mills term 
($\propto C_A^2$).
This amounts to a $30\%$ decrease of $a_2$ for the case of $n_f=0$,
and a $50\%$ decrease for $n_f=5$ (for $SU(3)$), which is the case 
needed for $t\bar t$ threshold investigations. 
This difference can be traced back to a specific set of diagrams,
as outlined below.

The origin of the discrepancy is Eq.~(14) in the second paper 
of \cite{Pe}. To explain the crucial point, let us introduce some 
notations first: We have two types of denominators,
$D_i\equiv D(k_i) = \frac1{k_i^2}\,$, stemming from gluon, ghost
and fermion propagators, and $S_i\equiv S(k_i) = 
\frac1{v\cdot k_i+i\ve}\,$, with $v=(1,\vec 0)$, stemming from the
source propagators.
The loop momenta are $k_1$, $k_2$, $k_3=k_1-q$, $k_4=k_2-q$,
$k_5=k_1-k_2$, where $q=(0,\vec q)$ is the external momentum.
We abbreviate the integration measure as 
$\int_i \equiv \mu^\e \int\!\frac{d^Dk_i}{(2\p)^D}\,$, while
products of propagators will be written like $D_1D_2\equiv D_{12}$
etc.

Adding the diagrams in question gives (neglecting the color
factors)
\def\erstes{\;\parbox{8mm}{\setlength{\unitlength}{1mm}
\begin{picture}(8,6) \thicklines
\put(0,0){\line(1,0){8}}
\put(0,6){\line(1,0){8}}
\put(0,0){\line(5,6){5}}
\put(3,0){\line(5,6){5}}
\put(0,6){\line(4,-3){8}}
\end{picture}}\;}
\def\zweites{\;\parbox{8mm}{\setlength{\unitlength}{1mm}
\begin{picture}(8,6) \thicklines
\put(0,0){\line(1,0){8}}
\put(0,6){\line(1,0){8}}
\put(0,6){\line(5,-6){5}}
\put(3,6){\line(5,-6){5}}
\put(0,0){\line(4,3){8}}
\end{picture}}\;}
\def\drittes{\;\parbox{8mm}{\setlength{\unitlength}{1mm}
\begin{picture}(8,6) \thicklines
\put(0,0){\line(1,0){8}}
\put(0,6){\line(1,0){8}}
\put(4,0){\line(0,1){6}}
\put(0,0){\line(4,3){8}}
\put(0,6){\line(4,-3){8}}
\end{picture}}\;}
\bea \label{X}
  \erstes + \zweites +\drittes &=& \int_1\int_2 \( 
  D_{235}S_{1125} + D_{145}S_{1125} + D_{235}S_{1122} \) \nonumber\\
  &=& \int_1\int_2 \( D_{145}S_{1125} + D_{235}S_{1225} \) \nonumber\\
  &=& \int_1\int_2 D_{145}S_{112} \( S_5+S_{\bar5} \) \;,
\eea
where the identity $S_1S_2=S_5(S_2-S_1)$
(compare \cite{Pe}, \para 4) was used for the last term of
the first line, and the trivial exchange of loop variables
$k_1 \leftrightarrow k_2$ was done in the last term of line two.
In the last line, $S_{\bar5}=S(-k_5)$. 
One then obtains
\bea
  S_5+S_{\bar5} &=& \frac1{(k_{10}-k_{20})+i\ve} + 
  \frac1{-(k_{10}-k_{20})+i\ve} \nonumber\\
  &=& -\frac{2i\ve}{(k_{10}-k_{20})^2+\ve^2} 
  \quad\stackrel{\ve\rightarrow0^+}{\longrightarrow}\quad -2\p i
  \d\left(k_{10}-k_{20}\right) \;.
\eea
Hence, contrary to the assumption in \cite{Pe}, the sum of
the integrands in Eq.~(\ref{X}) reduces to a delta distribution
multiplying the remaining propagators.

Now, considering the color traces as well as the 
gluon-source couplings, one gets as a contribution to 
the bare static potential (for simplicity, we use the
Feynman gauge here to make the point clear)
\bea
  {\rm diag.a6 + diag.b3} &=& -\frac{g^6}4\,C_FC_A^2 
  \(\erstes+\zweites\) 
  -\frac{g^6}2\,C_FC_A^2 \drittes 
  \nonumber\\
  &=& -\frac{g^6}4\,C_FC_A^2 \drittes -\frac{g^6}4\,C_FC_A^2 
  \(\erstes+\zweites+\drittes\) \nonumber \;.
\eea
While in \cite{Pe} the latter term was discarded, 
we evaluate it in $D=4-\e$ dimensions to give
\bea
  \erstes+\zweites+\drittes &=&
  \frac{6(D-4)(3D-11)}{(D-5)}\, \frac1{\vec q^2} 
  \int_1\int_2 D_{235} S_{12} \nonumber\\
  &=& -\frac1{32\p^2}\,\frac1{\vec q^2} +O(\e) \;.
\eea
Note that the factor of $(D-4)$ in the numerator cancels 
the single pole in the
scalar integral, such that only the constant
part of $a_2$ is affected by this discussion, while the pole terms
are not changed by the omission of this term. 
Hence, dividing out the overall factor 
$\(-\frac{C_F g^6}{(16\p^2)^2 \vec q^2}\)$, 
we identify the $2\p^2 C_A^2$ difference with respect to 
\cite{Pe}~\footnote{We thank M.~Peter for checking this result.}.

\newcommand{\sv}{{\scriptscriptstyle V}}
The static potential can be used for a definition of an
effective charge, which is conventionally called $a_\sv$.
Defining $V(\vec q^2)=-C_F 16\p^2 a_\sv/\vec q^2$,
one can use the knowledge of the three--loop coefficient $\b_2^\msbb$
\cite{ren} to derive the corresponding coefficient in the 
$V$--scheme from Eq.~(\ref{Result}). 
While $\b_0$ and $\b_1$ are universal, one finds
\bea
 \b_2^\sv &=& \b_2^\msbb - a_1\b_1 + (a_2-a_1^2)\b_0\\
 &=& \( \frac{206}3 + \frac{44\p^2}3 - \frac{11\p^4}{12} + 
 \frac{242}9\,\z(3) \)C_A^3 \nonumber\\
 && - \( \frac{445}9 + \frac{16\p^2}3 - \frac{\p^4}{3} + 
 \frac{704}9\,\z(3) \)C_A^2T_Fn_f 
 + \( \frac29 + \frac{224}9\,\z(3) \) C_AT_F^2n_f^2 \nonumber\\
 && - \( \frac{686}9 - \frac{176}3\,\z(3)  \)C_AC_FT_Fn_f
 + 2 C_F^2T_Fn_f +\(\frac{184}9 -\frac{64}3\,\z(3)\)C_FT_F^2n_f^2 \;.
\eea
The new value for $a_2$ leads, for $SU(3)$ and $n_f=5$, to 
a 50\% decrease of $\b_2^\sv$ compared to the formula 
given in \cite{Pe}.

Summarizing, we have re-calculated the two--loop static potential
by a method complementary to the approach in \cite{Pe}.
We have developed an algorithm which enables us to
work in general covariant gauges throughout. Confirming the
fermionic contributions to the two--loop coefficient $a_2$,
we find a substantial deviation in the pure gluonic part of $a_2$.
The source of the discrepancy could be identified.

We would like to thank W.~Buchm\"uller, M.~Spira, T.~Teubner and 
M.~Peter for valuable discussions and correspondence, respectively.

\end{document}